\documentstyle[fleqn]{tp}

\input psfig

\def\etal{{\it et\thinspace al.\/}}

\def\eg{{e.g}}

\def\bm{{\bf m}}
\def\bd{{\bf d}}

\def\hmpc{\,{\rm h^{-1}Mpc}}
\def\ihmpc{\, h\, {\rm Mpc^{-1}}}

\def\3hmpc{\, ( h^{-1} {\rm Mpc})^3}
\def\ln{{\rm ln}}

\def\oml{\Omega_\Lambda}

\def\lcdm{$\Lambda$CDM} 
\def\L{{\cal L}} 
\def\P{{\cal P}}

\begin{document}

\twocolumn[
\title{LARGE-SCALE MASS POWER SPECTRUM FROM \\ PECULIAR VELOCITIES}
\author{Idit Zehavi$^{1,2}$\\ 
{\it $^1$Racah Institute of Physics, The Hebrew University,}\\
{\it Jerusalem 91904, Israel} \\
{\it $^2$NASA/Fermilab Astrophysics Group, Fermi National Accelerator}\\ 
{\it Laboratory, Box 500, Batavia, IL 60510-0500, USA}}
\vspace*{16pt}   

ABSTRACT.\
The power spectrum of mass density fluctuations is estimated from 
the Mark~III and the SFI catalogs of peculiar velocities by
applying a maximum likelihood analysis, using parametric models
for the power spectrum and for the errors. Generalized CDM models 
with and without COBE normalization are used. The applications to 
the two different data sets give consistent results. The general 
result is a relatively high amplitude of the power spectrum, \eg, at
$k=0.1 \,h\, {\rm Mpc^{-1}}$ we find $P(k) \Omega^{1.2} = (4.5\pm2.0)
\times 10^3 \,(h^{-1} {\rm Mpc})^3$, corresponding to
$\sigma_8 \Omega^{0.6} = 0.85 \pm 0.2$. Model-dependent 
constraints on combinations of cosmological parameters are obtained 
for families of COBE-normalized CDM models. These can roughly be 
approximated by $\Omega\, {h_{60}}^\mu\, n^\nu = 0.6 \pm 0.2$, where 
$\mu = 1.3$ and $\nu = 3.7,\ 2.0$ for flat \lcdm\, models with and without
tensor fluctuations respectively. For open CDM, without tensor fluctuations, 
the powers are $\mu = 0.9$ and $\nu = 1.4$. The quoted error-bars reflect 
the $90\%$ formal likelihood uncertainty for each model and the variance 
among different models and between catalogs. This is a brief review of 
a collaborative project (for more details, see Zaroubi \etal\ 1997, 
Freudling \etal\ 1998). Preliminary constraints in the $\Omega-\oml$ plane
are presented as well.
\endabstract]

\markboth{Idit Zehavi}{Large-Scale Mass Power Spectrum from Peculiar
Velocities}

\small

\section{Introduction}
\label{sec:intro}
In the standard picture of cosmology, structure evolved from small
density fluctuations that grew by gravitational instability. These
initial fluctuations are assumed to have a Gaussian distribution
characterized by the power spectrum (PS). On large scales, the 
fluctuations are linear even at late times and still governed by the 
initial PS. The PS is thus a useful statistic for large-scale structure, 
providing constraints on cosmology and theories of structure formation. 

The galaxy PS has been estimated in recent years from several redshift
surveys (see reviews by Strauss \& Willick 1995; Strauss 1998). 
Alternatively, one can estimate the PS using measurements of peculiar 
velocities, which are directly related to the {\it mass} density 
fluctuations. Velocities are also sensitive to larger scales and thus
subject to weaker non-linear effects. 
In this work, we develop and apply a likelihood analysis (first proposed 
by Kaiser 1988) in order to estimate the mass PS from peculiar velocity
catalogs. The method, acting on the ``raw'' peculiar velocities without 
additional processing, utilizes much of the information content of the 
data. It takes into account properly the measurement errors and the 
finite discrete sampling. The simplifying assumptions made are that the 
velocities follow a Gaussian distribution and that their correlations can 
be derived from the density PS using linear theory. 

Two catalogs are used for this purpose. One is the Mark~III catalog of
peculiar velocities, a compilation of several data sets, consisting of 
roughly 3000 spiral and elliptical galaxies within a volume of 
$\sim 80 \hmpc$ around the local group, grouped into $\sim 1200$ objects 
(Willick \etal\ 1995, 1996 1997). The other is the recently completed SFI 
catalog, a homogeneously selected sample of $\sim 1300$ spiral field 
galaxies, designed to minimize effects of combining disparate data sets 
(Haynes \etal\ 1998; Wegner \etal\ 1998).  In both catalogs, the typical 
relative distance errors of individual galaxies are $15-20\%$, and both 
data sets are carefully corrected for the various systematic biases. 
It is interesting to compare the results of the two catalogs, especially 
in view of apparent discrepancies in the appearance of the velocity fields 
(\eg, da Costa \etal\ 1996, 1998). 

\section{Method}
\label{sec:method}
Given a data set $\bd$, the goal is to estimate the most likely model
$\bm$. Invoking a Bayesian approach (and assuming a uniform prior), this 
can be turned to maximizing
the likelihood function $\L \equiv \P(\bd|\bm)$, the probability of
the data given the model, as a function of the model parameters.
Under the assumption that both the underlying velocities and the
observational errors are Gaussian random fields, the likelihood
function can be written as 
\begin{eqnarray*} 
{\cal L} & = & [ (2\pi)^N \det(R)]^{-1/2} \\
&& \quad \times 
\exp\left( -{1\over 2}\sum_{i,j}^N {d_i R_{ij}^{-1} d_j}\right), 
\end{eqnarray*}
where $\{d_i\}_{i=1}^{N}$ is the set of observed peculiar
velocities and $R$ is their correlation matrix. $R$ involves the theoretical 
correlation, calculated in linear theory for each assumed cosmological
model, and the estimated covariance of the errors.

The likelihood analysis is performed by choosing some parametric
functional form for the PS. 
Going over the parameter space and calculating the likelihoods for the
different models, one finds the PS parameters for which the maximum 
likelihood is obtained.  Confidence levels are
estimated by approximating $-2\ln\L$ as a $\chi^2$ distribution with
respect to the model parameters. Note that this method, based on
peculiar velocities, essentially measures $f(\Omega)^2P(k)$ and not
the mass density PS by itself. We extensively test the method 
using realistic mock catalogs, designed to mimic in detail the real 
catalogs (Kolatt \etal\ 1996; Eldar \etal\ 1998).

We use several models for the PS. One of these is the so-called
$\Gamma$ model, where we vary the amplitude and the shape-parameter
$\Gamma$.  The main analysis is done with a suit of generalized CDM
models, normalized by the COBE 4-year data. These include open models,
flat models with a cosmological constant and tilted models with or
without a tensor component.  The free parameters are then the 
mass-density parameter $\Omega$, the Hubble parameter $h$
and the power index $n$.

Here, as in any method for estimating the PS, the recovered PS is 
sensitive to the assumed observational errors, that enter as well
the correlation matrix $R$. To alleviate this problem, we extend the 
method such that also the magnitude of these errors is determined by the 
likelihood analysis. This is done by adding free parameters that 
govern a global change of the assumed errors, in addition to modeling 
the PS, and provides some reliability check of the magnitude of the 
errors. We find, for both catalogs, a good agreement with the original 
error estimates, thus allowing for a more reliable recovery of the PS.

\section{Results}
\label{sec:res}
\begin{figure*}
\vskip -2cm
\centering\mbox{\psfig{figure=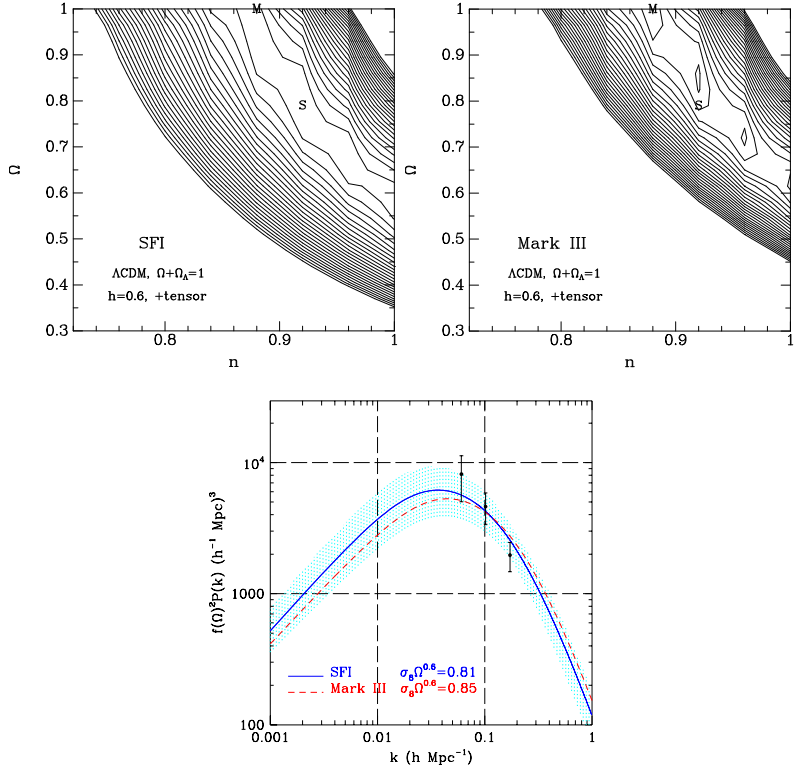,height=15.5cm}}
\vskip -3cm
\caption[]{Likelihood analysis results for the COBE-normalized flat
\lcdm\ model with $h=0.6$. Shown are $\ln\L$ contours in the $\Omega-n$ 
plane for SFI (top left panel) and for Mark~III (top right). The best-fit 
parameters for SFI and Mark~III are marked, on both, by `S' and `M' 
respectively. The lower panel shows the corresponding maximum-likelihood 
PS for SFI (solid line) and for Mark~III (dashed). The shaded region is 
the SFI $90\%$ confidence region. The three solid dots mark the PS calculated 
from Mark~III by Kolatt and Dekel (1997), together with their quoted  
$1\sigma$ error-bar.}
\label{fig:sfi_mark}
\end{figure*}

Figure~\ref{fig:sfi_mark} shows, as a typical example, the results for 
the COBE-normalized flat \lcdm\ family of models, with a tensor component 
in the initial
fluctuations, when setting $h=0.6$ and varying $\Omega$ and $n$. 
Shown are $\ln\L$ contours for the SFI catalog and for Mark~III. As 
can be seen from the elongated contours, what is determined well is not 
a specific point but a high likelihood ridge, constraining a degenerate 
combination of the parameters roughly of the form $\Omega\, n^{3.7} = 
0.59 \pm 0.08$, in this case. The corresponding best-fit PS for the two 
catalogs is presented as well, with the shaded region illustrating the 
$90\%$ confidence region obtained from the SFI high-likelihood ridge.

These results are representative for all other PS models we tried. For
each catalog, the different models yield similar best-fit PS, falling
well within each others formal uncertainties and agreeing especially
well on intermediate scales ($k \sim 0.1\ihmpc$). The similarity of the 
PS obtained from SFI with that of Mark~III, which is seen in the figure, 
is illustrative of the other models as well. This indicates that the
peculiar velocities of the two catalogs, with their respective error 
estimates, are consistent with arising from the same underlying mass 
density PS. This does not preclude possible differences 
that are not picked up by this statistic, but can be viewed as another
indication of the robustness of the results. Note also the agreement 
with an independent measure of the PS from the Mark~III catalog, using 
the smoothed density field recovered by POTENT (the three dots; Kolatt 
\& Dekel 1997).

\begin{figure*}
\vskip -0.5cm
\hskip -0.5cm 
\centering \mbox{\psfig{figure=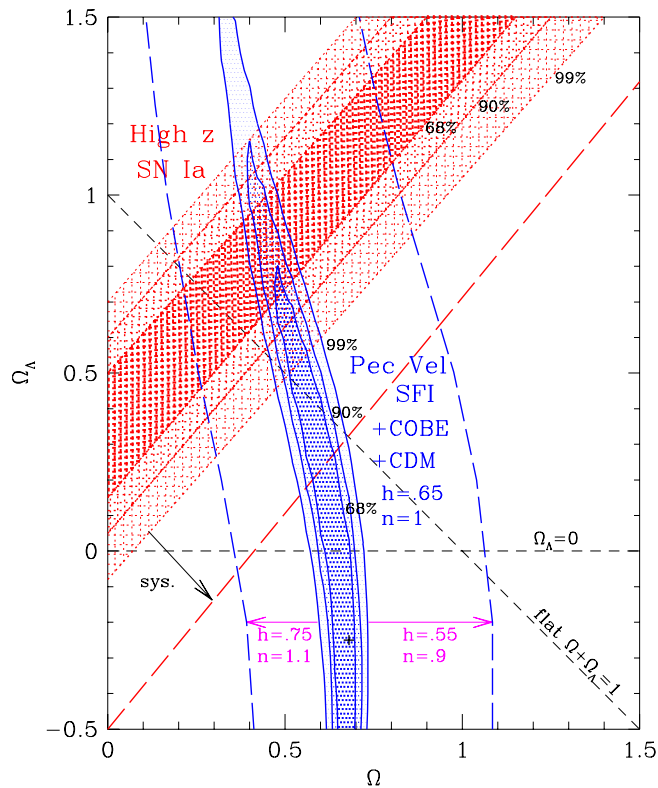,height=14cm}}
\vskip -3cm
\caption[]{Constraints in the $\Omega - \oml$ plane coming from a 
likelihood analysis of the SFI peculiar velocities (the approx. vertical
contours) and from the high-redshift SNe Ia (the diagonal contours; 
Perlmutter \etal\ 1998). The $68$, $90$ and $99\%$ confidence regions are
shown for both. The estimated worst-case systematic error of the SNe results 
is denoted by the corresponding dashed line. The SFI likelihood contours are 
for the case of $n=1$, $h=0.65$. The shifted dashed lines illustrate the 
estimated effect of changing the values of these parameters (Zehavi \& Dekel
1998).  } 
\label{fig:om_lam}
\end{figure*}

The robust result, for both catalogs and all models, is a relatively
high PS, with $P(k) \Omega^{1.2} = (4.5\pm2.0)\times 10^3 \3hmpc$ at
$k=0.1\ihmpc$.  An extrapolation to smaller scales using the different
CDM models gives $\sigma_8 \Omega^{0.6} = 0.85 \pm 0.2$. (Such values 
are also obtained when assuming for the PS the $\Gamma$ model or the 
generalized CDM models with a {\it free} amplitude.) The high-likelihood 
ridge is a feature of all COBE-normalized CDM models. The general 
constraints on the combination of cosmological parameters
is of the sort $\Omega\, {h_{60}}^\mu\, n^\nu = 0.6 \pm 0.2$, where
$\mu = 1.3$ and $\nu = 3.7,\ 2.0$ for flat \lcdm\, models with and without
tensor fluctuations respectively. For open CDM, without tensor
fluctuations, the powers are $\mu = 0.9$ and $\nu = 1.4$.  For the
span of models checked, the PS peak is in the range $0.02\leq k\leq
0.06\ihmpc$. The shape parameter of the $\Gamma$ model is only weakly
constrained to $\Gamma = 0.4\pm 0.2$. 
These error-bars are crude, reflecting the $90\%$ formal likelihood
uncertainty for each model, the variance among different models and
between catalogs. 
Care should also be given to possible systematics that could still 
plague the results, arising perhaps from non-linear effects or some
peculiarities in the data (Freudling \etal\ 1998).

\section{Further Analysis}
\label{sec:om_lam}

We have recently extended the analysis of COBE-normalized CDM models
to models with general values of $\Omega$ and $\oml$ (Zehavi \& Dekel 
1998). Although $\oml$ comes in only indirectly through the COBE 
normalization, such results are particularly interesting as they can 
be compared to other constraints in the $\Omega - \oml$ plane, such as 
the recent results coming from high-redshift type Ia supernovae (SNe Ia; 
Perlmutter \etal\ 1998; Riess \etal\ 1998).

Figure~\ref{fig:om_lam} illustrates such  constraints in the $\Omega-\oml$ 
plane, showing $\ln\L$ contours for the SFI catalog, for fixed values of 
$n=1$ and $h=0.65$. The peculiar velocity analysis appears to constrain 
an elongated ridge in this plane of a nearly fixed $\Omega$ and varying 
$\oml$. As demonstrated in the plot, a change in the values of $n$ and $h$ 
essentially shifts the ridge toward either a higher or lower $\Omega$, for 
smaller and larger values of these parameters, respectively. This is another 
manifestation of the degeneracy between these parameters mentioned earlier. 
The acceptable range of these parameters is therefore needed to be determined 
by other external constraints. 
The results of the same analysis applied to the Mark~III catalog are
here as well fairly similar to the SFI ones, except for a slightly 
stronger preference toward smaller values of $\oml$.

The confidence contours in this parameter plane obtained by the Supernova 
Cosmology Project (Perlmutter \etal\ 1998) are sketched as well in 
Figure~\ref{fig:om_lam}. (These results are consistent with the findings 
of the High-z Supernova Search Team results, Riess \etal\ 1998.)
Taking into consideration concurrently these two independent sets of 
constraints seem to imply a considerable contribution from both $\Omega$ 
and $\oml$. While the range of models consistent with the high-z SNe 
findings alone includes low $\Omega$ $+$ low $\oml$ models (and even 
$\oml < 0)$, the peculiar velocities analysis appears to rule out these 
models, and taken together the two make a stronger case for a positive 
cosmological constant.

Work in progress includes an attempt to merge the Mark~III and SFI 
catalogs and perform a joint likelihood analysis, looking specifically 
at the cross-correlations of the two data sets, that may entail valuable 
information.
Another aim is to perform a similar analysis on other velocity data, such 
as velocities of galaxy clusters (\eg, Smith \etal, these proceedings) or 
SNe Ia velocities, which probe out to larger scales with relatively high 
accuracy.
Lastly, an interesting prospect is to do a simultaneous analysis of
velocity data together with other kinds of data, like redshift surveys
and CMB experiments (see Webster \etal\ 1998; Lahav \& Bridle, these 
proceedings). The distinct types of data complement one another, each 
constraining different combinations of the cosmological parameters, and 
together may remove the degeneracy and set tight constraints.

\section*{Acknowledgments}
I thank my collaborators in this work L.N.\ da Costa, A.\ Dekel, A.\ Eldar, 
W.\ Freudling, R.\ Giovanelli, M.P.\ Haynes, Y.\ Hoffman, T.\ Kolatt, 
J.J.\ Salzer, G.\ Wegner and S.\ Zaroubi. This work was supported by the 
DOE and the NASA grant NAG 5-7092 at Fermilab.
 

\end{document}